# Keeping Exoplanet Science Caffeinated with ESPRESSO

Louise Dyregaard Nielsen[1]
Julia Victoria Seidel[1]

[1] ESO

The ESPRESSO spectrograph at ESO's Very Large Telescope (VLT) has, since it began science operations in October 2018, revolutionised exoplanet science. The combination of the large VLT mirrors and the high resolution and stability of the spectrograph is enabling the detection of small, low-mass planets as well as detailed studies of the planets' atmospheres. In this article we present a brief overview of the first results from ESPRESSO and a hopeful glimpse towards the ultimate goal of reaching the radial velocity precision of 10 cm s$^{-1}$ needed to detect an Earth-like planet.

## The quest for temperate, rocky planets

Is life a common constituent of the Universe? This overarching question is currently driving the construction of telescopes (small, extremely large and space-based alike) and the development of new instruments. We know that planets themselves are very common; at least 50% of Sun-like stars have one planet, possibly more, around them (Fressin et al., 2013). Rocky planets seem to be one of the most predominant types of exoplanets out there, 10 times more frequent than giant gas planets when considering close-in orbits. However, when it comes to determining the occurrence of Earth-sized planets we are limited by current survey sensitivity, especially when probing orbital periods beyond 100 days (Fulton & Petigura, 2018).

The radial velocity technique has, since the first discovery of an exoplanet around a main sequence star by Mayor & Queloz (1995), proven useful for discovering new exoplanets and measuring the masses of known transiting exoplanets. The wobble of a host star, induced by the gravitational pull of an orbiting planet, is translated into a periodic shift of the stellar absorption lines as the stellar light is red- and blue-shifted away from us and towards us, respectively, as seen in Figure 1. This subtle effect can be measured by high-resolution spectroscopy with stabilised instruments and reliable wavelength calibrations.

As new generations of instruments have entered the stage, astronomers have been able to push towards less and less massive planets. Figure 2 shows the masses of known exoplanets measured with the radial velocity technique (in Earth-masses; note the logarithmic scale) as a function of their discovery year. For systems where the orbital inclination is not known, the projected mass is used. Over the last 15 years, an almost linear evolution toward less massive planets has delivered the first Earth-mass exoplanets.

## Extremely precise radial velocities with ESPRESSO

After the start of operations at ESO's Very Large Telescope (VLT) in October 2018, and a subsequent fibre upgrade in June 2019, the Echelle SPectrograph for Rocky Planet and Stable Spectroscopic Observations (ESPRESSO; Pepe et al., 2021) has started an era of extremely precise radial velocity measurements. Compared to its predecessor, the High Accuracy Radial velocity Planet Searcher (HARPS) on the ESO 3.6-metre telescope (Mayor et al., 2003), ESPRESSO benefits from a larger collecting area, wider wavelength range and better instrument stability and calibration. The instrument can be fed with light from any of the Unit Telescopes (UTs) at Paranal Observatory, or from all of them to mimic an effective collecting area corresponding to a 16-metre telescope. The 4-UT mode is offered with extragalactic astronomy also in mind, though in this article we will focus on the advances in exoplanet science only.

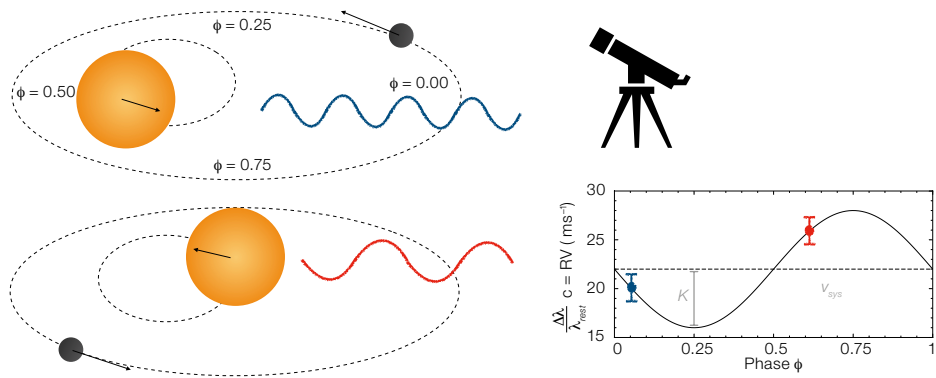

Figure 1. As a planet and star are orbiting their common centre of mass, the stellar light is red- and blue-shifted. The semi-amplitude of the stellar radial velocity shift, $K_{RV}$, is directly related to the mass ratio of planet and star, and the orbital period.

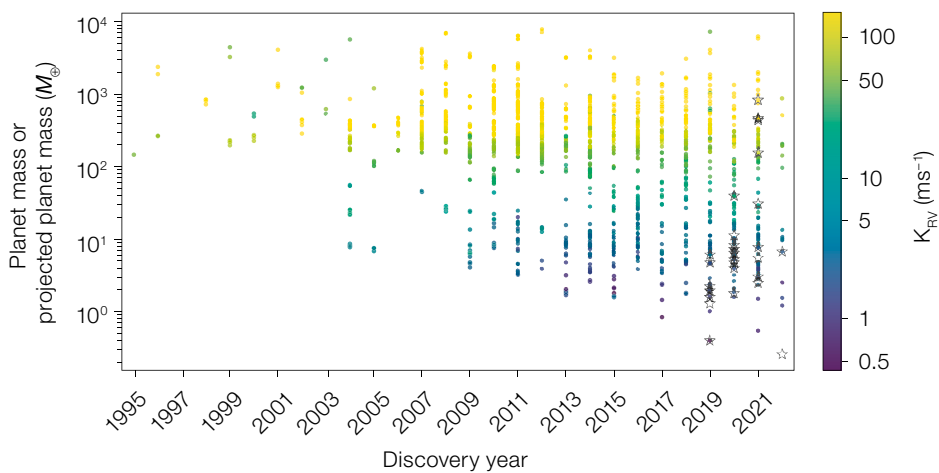

Figure 2. Exoplanet masses, or projected masses, based on radial velocity measurements versus the year of discovery. Note that the y-axis is in Earth-masses and is logarithmic. The colour scale is radial velocity semi-amplitude, $K_{RV}$. Masses determined with ESPRESSO are highlighted with stars, and empty stars are for planets not yet confirmed. Based on data from NASA Exoplanet Archive[1], April 2022.

## Unveiling the true nature of our closest neighbour

The nearest star to our Solar System, Proxima Centauri, has been monitored closely with ESPRESSO as part of an effort to discover low-mass planets in the habitable zones around nearby stars (Hojjatpanah et al., 2019). Proxima Centauri was known to host an Earth-mass planet, known as Proxima b, in an 11.2-day orbit (Anglada-Escudé et al., 2016) as well as a seven times more massive candidate on a 5-year orbit (Damasso et al., 2020). Using ESPRESSO, Suárez Mascareño et al. (2020) independently confirmed the orbit of Proxima b at 11.2 days, while also presenting hints of a third candidate in the system. This planet candidate was recently established by Faria et al. (2022) after collecting an additional season of ESPRESSO data. Based on three seasons of ESPRESSO data, Faria et al. (2022) find a minimum mass of 0.25 that of Earth for this close-in candidate, which is orbiting at a distance of just 0.029 astronomical units from its host star. The radial velocity data are shown in Figure 3, where the phase-folded data for the two planets with the best fit model are shown. The RMS of the residuals across the three seasons of observations is just 0.26 ms$^{-1}$.

With the discovery of a third rocky planet candidate around Proxima Centauri, our closest stellar system illustrates the potential for the discovery of low-mass planets around nearby stars. Given the low inherent luminosity of the host star, the three planets orbit just inward, within, and outward of the zone where liquid water could exist, commonly refereed to as the potential habitable, or goldilocks, zone.

When analysing the radial velocity observations from ESPRESSO and other spectrographs, one of the biggest challenges is separating the imprint of stellar activity from the signals caused by planets. The presence of active regions (for example, starspots) at the stellar surface creates radial velocity variations that can mimic a planet. This has been tackled using statistical methods combined with so-called activity indicators, which are auxiliary observations that measure the level of stellar activity (Pont, Aigrain & Zucker, 2011; Dumusque et al., 2017; Zicher et al., 2022). Furthermore, the large wavelength coverage and high radial velocity precision of ESPRESSO have made it possible to verify that signals are constant across the wavelength range (as planetary signals ought to be) and thereby to separate them from signals due to stellar activity.

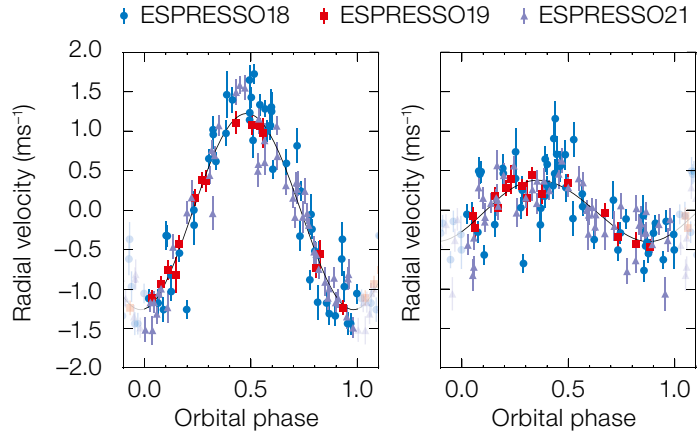

Figure 3. Radial velocity measurements of Proxima Cen from Faria et al. (2022). The two panels show the data phase folded on the epoch and period of the two detected planets along with the best fit model. The RMS of the residuals is just 0.26 ms$^{-1}$.

## Measuring the masses of transiting planets

Transiting exoplanets present unique opportunities for in-depth characterisation as we can potentially measure both their radius and mass as well as constrain full 3D orbits and determine atmospheric properties. The L 98-59 system contains three transiting planets (Kostov et al., 2019; Cloutier et al., 2019) as well as two additional candidates seen only in radial velocity measurements with ESPRESSO (Demangeon et al., 2021). L 98-59 b is a rocky planet with half the mass of Venus and is currently the lowest-mass planet measured using radial velocities.

Another interesting system for detailed studies is WASP-47, which has three inner transiting planets: a super-Earth at 0.79 days, a hot Jupiter at 4.16 days and a Neptune at 9.03 days (Hellier et al., 2012; Becker et al., 2015). Additionally, WASP-47 is known to host an outer giant planet with an orbital period of 1.6 years (Neveu-VanMalle et al., 2016), though it is uncertain whether it is in a transiting orbital configuration as seen from Earth. Recently, Bryant & Bayliss (2022) revisited the WASP-47 system using ESPRESSO and space-based photometry. Thanks to a high-cadence observing strategy they were able to refine the mass of WASP-47 e, the smallest, innermost planet in the system. Compared to the ensemble of known super-Earths, the density of WASP-47 e is found to be low, which could be caused by tidal interaction with the hot Jupiter in the system.

## Probing exoplanet atmospheres with resolved spectral lines

In the few years that ESPRESSO has given us high-resolution data, it has also proven to be a powerful tool for characterising exoplanet atmospheres. With its high-fidelity spectra, many approaches to studying exoplanet systems were automated and various codes are now available to understand the 3D orbits and atmospheric constituents of exoplanets. Some examples are the CaRM code to learn more about the Rossiter-McLaughlin effect (Cristo et al., 2022), CHOCOLATE — a new chromatic Doppler tomography technique (Esparza-Borges et al., 2022), and the RRM revolutions approach (Bourrier et al., 2022) on GJ436 b, which lets us derive system parameters all the way down in size to Neptunes.

With all these new techniques, it is only a question of time before controversy arises. But who would have thought that the first bombshell dropping from ESPRESSO data would be related to the all-time favourite exoplanet, HD209458 b? Claimed as the first detected exoplanet atmosphere by Charbonneau et al. (2002), it has since been the poster child of atmospheric detections, with currently more than 75 peer-reviewed articles to its name. Yet, analysing ESPRESSO data taken over two individual transits revealed that the detections previously claimed for sodium, potassium, magnesium, iron and a handful of other molecules and atoms can be attributed solely to the distortion of the stellar lines by the Rossiter-McLaughlin effect (Casasayas-Barris et al., 2021).

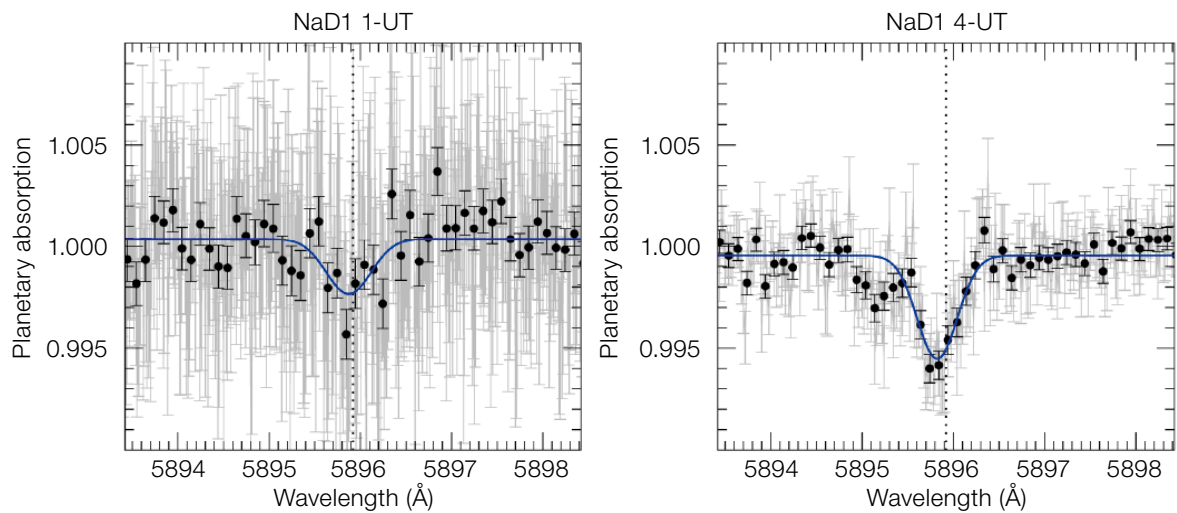

Figure 4. Sodium detection in WASP-121 b with ESPRESSO with one UT on the left and four UTs on the right. From Borsa et al. (2021).

While not as emotionally charged as the results for HD209458 b, ESPRESSO has also contributed to the understanding of the sub-Saturn WASP-127 b and the hot Jupiter WASP-19 b. For WASP-127 b, two independent studies of HARPS data came to different conclusions about the sodium detection strength. Allart et al. (2020) ultimately demonstrated with ESPRESSO data that WASP-127 b has a shallow sodium feature. Similarly, Sedaghati et al. (2021) added useful information to the ongoing study of WASP-19 b by providing various non-detections (for example, of Fe) and solidifying the detection of TiO, thus resolving the differences in the results obtained with FORS (Sedaghati et al., 2017) and IMACS (Espinoza et al., 2019).

Other atmospheric detections include sodium and potassium for the hot Jupiter WASP-117 b (Carone et al., 2021) and the confirmation of atmospheric sodium for the Neptune desert planet WASP-166 b (Seidel et al., 2022 and papers to follow).

But one of the first detection results from ESPRESSO directly turned out to be spectacular: Tabernero et al. (2021) re-observed the ultra-hot Jupiter WASP-76 b with ESPRESSO and confirmed the extremely broadened sodium signal as well as various other detections, starting off the race for ESPRESSO data on exoplanet atmospheres.

## From static to time-resolved, from upper-limits to precise values: the ESPRESSO effect

WASP-76 b, the first ultra-hot Jupiter with a resolved sodium signature, provided the community with the most precise high-resolution dataset of any exoplanet atmosphere to date. The same dataset was analysed by Ehrenreich et al. (2020) who go beyond the integrated detection of elements. Instead of just claiming iron in its atmosphere, they trace iron as a function of phase, providing a time-resolved map of iron across the terminator of WASP-76 b. This leads them to a direct, time-resolved detection of a 5-km s$^{-1}$ dayside-to-nightside wind in the lower atmosphere — a first in our understanding of exoplanet atmospheric dynamics. On the integrated transmission spectrum, the resolved sodium doublet opens the observational window on the exoplanet atmosphere to include even higher layers, probing all the way to the thermosphere. The same ESPRESSO dataset on WASP-76 b allowed the direct retrieval of 3D wind patterns by Seidel et al. (2021), who found exactly the same speeds for a dayside-to-nightside wind in the lower atmosphere as Ehrenreich et al. (2020) had done, confirming their observational results. Seidel et al. (2021) also find a vertical wind connecting the lower atmosphere to the photoevaporation-driven mass loss in the exosphere.

The journey with ESPRESSO to a more detailed understanding of exoplanet atmospheres has, however, only just begun. The 4-UT mode of ESPRESSO has recently provided the sharpest line shapes we have ever seen for a plethora of spectral lines, opening up a whole new world of time-resolved observations (Borsa et al., 2021; see Figure 4).

### Acknowledgements

We thank João Faria for valuable comments and feedback on this article and for supplying the radial velocity figures for Proxima Centauri reproduced at Figure 3.

Links

[1] NASA Exoplanet Archive: https://exoplanetarchive.ipac.caltech.edu/

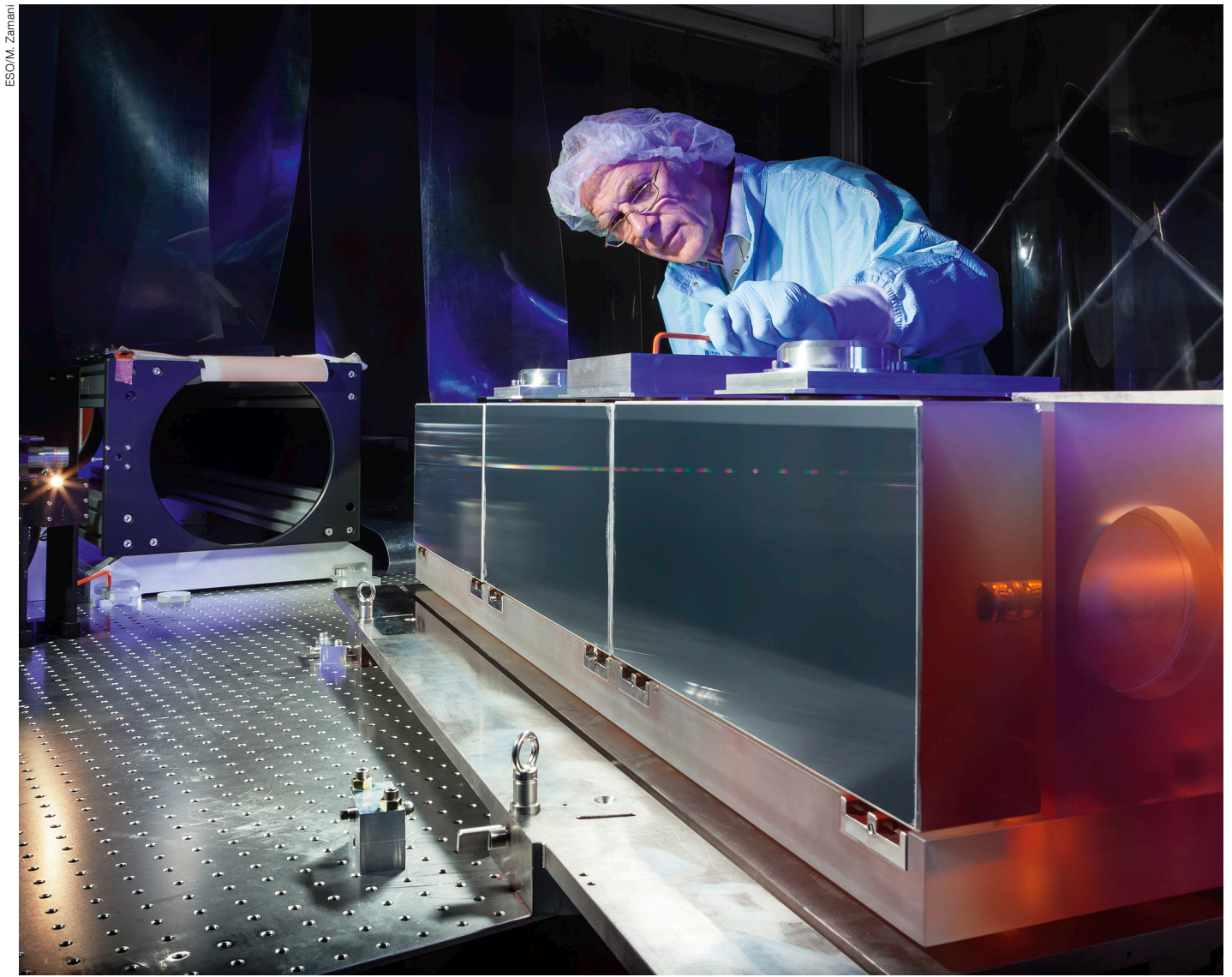

The huge diffraction grating at the heart of the ultra-precise ESPRESSO spectrograph — the next generation in exoplanet detection technology — is pictured undergoing testing in the cleanroom at ESO Headquarters in Garching bei München, Germany.